\documentclass{epl}

\title{Efficient loading and cooling in a dynamic optical evanescent-wave microtrap}
\shorttitle{Dynamic optical microtraps}
\author{Peter Domokos\thanks{On leave from the Research Institute for
Solid State Physics and Optics, Hungarian Academy of Sciences} \and
Helmut Ritsch\thanks{E-mail: \email{Helmut.Ritsch@uibk.ac.at}}}
\institute{Institut f{\"u}r Theoretische Physik, Universit{\"a}t Innsbruck - 
Technikerstr.\ 25, A-6020 Innsbruck, Austria}

\begin{document}

\maketitle

\begin{abstract}
We calculate the loading efficiency and cooling rates in a bichromatic
optical microtrap, where the optical potentials are generated by
evanescent waves of cavity fields at a dielectric--vacuum
interface. The cavity modified nonconservative dynamic light forces
lead to efficient loading of the atoms as well as cooling without the
need for spontaneous emission. Steady--state temperatures well below
the trap depth, reaching the motional quantum regime, yield very long
capturing times for a neutral atom.
\end{abstract}

Evanescent waves constitute an essential ingredient in various atom
optical elements, such as mirrors \cite{balykin88} for gravitational
atom optical cavities \cite{kasevich90,aminoff93} and for matter
waveguides \cite{barnett00}, or in optical surface traps
\cite{ovchinnikov97,gauck98}.  The electromagnetic radiation field
experiencing total internal reflection at a dielectric-vacuum
interface penetrates into the vacuum with an exponentially decaying
field amplitude. Large intensity gradient occurs close to the
surface. The resulting optical dipole force on neutral atoms is
unusually large and spatially well localized.

In extension of the simple repulsive geometry used for mirrors,
Ovchinnikov et al.\ \cite{ovchinnikov91} proposed a wavelength-size
optical microtrap based on the evanescent field of {\it two} radiation
modes with different spatial dependence. A strongly red detuned field
with a mode function proportional to $\exp(-k x)$ yields a long-range
attractive force towards the surface and a second short-range blue
detuned field decaying as $\exp(-2 k x)$ provides for a short-range
repulsive wall. The combination of these two fields generates a tiny
but deep trap with a Morse-type potential. Such a trap has a number of
remarkable features and potential applicabilities, e.g., it is
suitable for studying the interaction between cold atoms and a surface
\cite{gorlicki00}, or realizing 2D quantum gas. For a fixed given
field configuration the potential is conservative and one can expect
long trapping times. However, an efficient loading and cooling
mechanism is needed to fill the trap. In the case of the
gravito-optical surface trap
\cite{ovchinnikov97}, for example, optical pumping between hyperfine 
states is applied to implement a Sisyphus-type cooling on the bouncing
cesium atoms\cite{ovchinnikov95,desbiolles96}. The waveguide for argon
atoms in \cite{gauck98} also resorts to local optical pumping to
achieve an incoherent transfer to a trapped metastable state. In all
of these mechanisms spontaneous emission is in the core of the
cooling process. This has of course the immediate consequence of
heating by momentum diffusion and density limitations due to
reabsorption. Moreover, the extra lasers and optics needed for the
cooling pose severe restrictions in really microscopic setups.

In this Letter we investigate an efficient method for loading and
cooling atoms in a bichromatic evanescent-wave dipole trap. As a main
virtue of our scheme, there is no need of the atom to change its
internal state or undergo spontaneous emission.  The two evanescent
field modes, in addition to producing the optical potential trap,
simultaneously take over the role of a dissipation channel for the
kinetic energy of the atoms. The system relies on a {\it dynamic}
trapping field that imposes a nonconservative motion on an atom. A
central requirement to implement this scheme is a significant coupling
between the field modes and the atomic dipole. This implies a geometry
with the two fields being confined in small mode volumes, which can be
achieved 
with monolithic resonators (microspheres, microdisks, hollow fibers).

In the strong atom-field coupling regime the dynamical behaviour of
the system is rather complex. The atom moves guided by the dipole
force potential but at the same time acts as a moving refractive index
with a non-negligible effect on the field dynamics. As a consequence
the field amplitudes depend on the atomic position. A coupled dynamics
arises in which the atom can be damped via the cavity field decay
\cite{horak97,pinkse00,hood00}. 
The basic mechanism responsible for the kinetic energy
loss can be elucidated on the simple example of an atom colliding with
a single blue-detuned evanescent field at a given surface. For
suitable parameters, the photon numbers in the repulsive field may
vary such that the atom approaching the surface climbs a higher wall
than the one it is repelled from when moving away. Such a bounce can
be viewed as a Sisyphus-type process with photon number states
replacing the atomic hyperfine states. The only further necessary
ingredient in our system is now a second long-range attractive field
which pulls the atoms repeatedly towards the surface.  In principle
this system can form a 3D trap. For simplicity and in order to work
out the basic principles, we will, however, investigate a generic 1D
version of such a trap where both the attractive and repulsive fields
are dynamic.

The model system under consideration is schematically represented in
figure 1. We restrict the atomic motion to one spatial dimension $x$
perpendicular to the surface.  The incidence angles of the totally
reflected fields are adjusted such that the penetration depth of the
blue field ($b$) is just the half of that of the red one ($r$), with
frequencies properly chosen to match the specific atomic level
scheme. The field mode functions above the surface are given simply by
$f_r(x)=\exp(-k x)$ and $f_b(x)=\exp(-2 k x)$ for the modes $r$ and
$b$, respectively. 
The two modes are assumed to couple to different dipole transitions in
the atom. Their frequencies are sufficiently detuned from the atomic
transition to reduce atomic excitation and spontaneous emission. The
atom-field detuning is chosen $\Delta_A$ for both dipole transitions
but with opposite signs for the 'red' and 'blue' fields. The
frequencies of the actual red and blue fields are also detuned from
the cavity resonances. These pump-mode detunings are set to the same
negative value $\Delta_C$ many orders of magnitude less than
$\Delta_A$ (scheme in figure 1 is not to scale).
\begin{figure}
\oneimage[width=14cm]{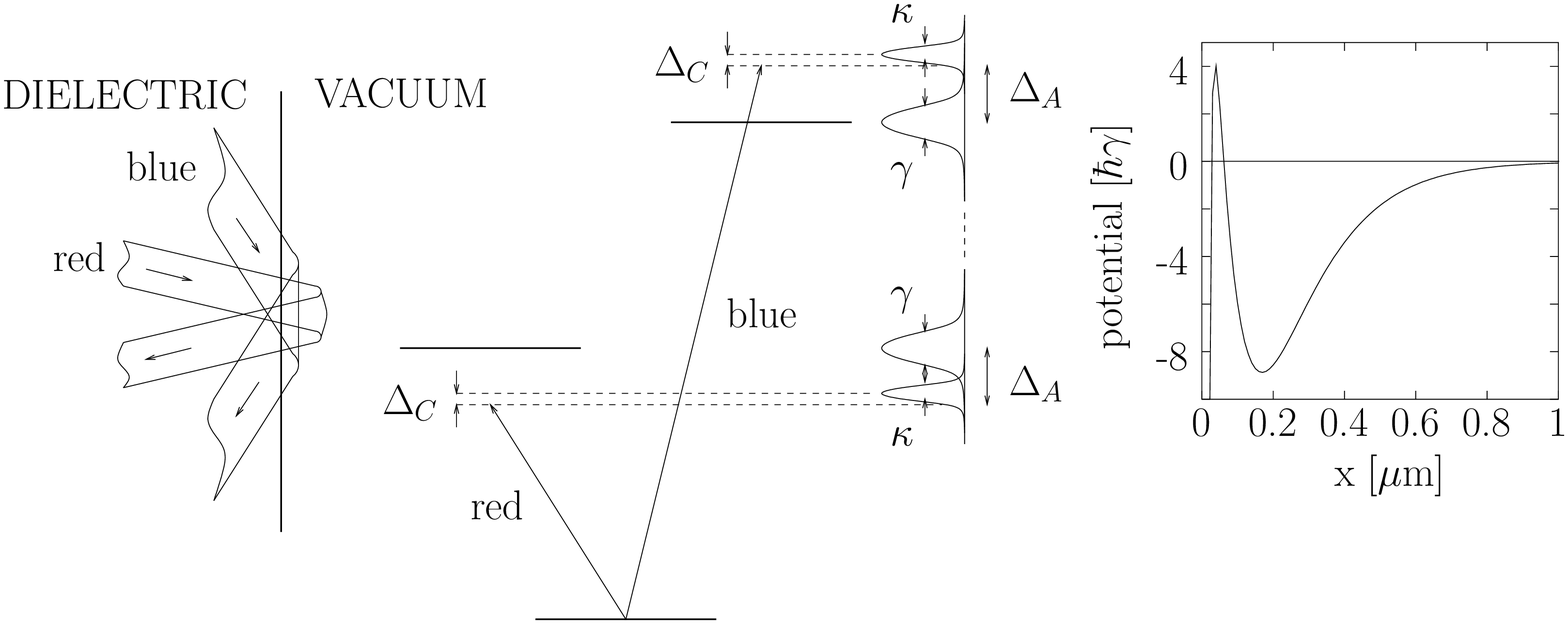}
\caption{Left: The geometrical arrangement of the beams. The
penetration depth of the blue is just the half of that of the red
field. Center: Excited transitions and detunings, not to scale,
involved in the scheme. Right: The adiabatic potential for the atomic
motion, including the strong Van der Waals attraction of the surface.}
\label{f.1}
\end{figure}
For the sake of simplicity, the mode linewidth $\kappa$ and the atomic
linewidth $\gamma$, as well as the dipole coupling constant $g$, are
assumed to be the same for both fields and we consider strong
coupling, i.e., $g$ being of the same order of magnitude as $\kappa$
and $\gamma$. This is about an order of magnitude less than it could
be in specific microsphere setups \cite{treussart94,mabuchi94} but
still fairly large. Although a higher $g$ would improve the system
performance, we aim to demonstrate that the actual ``top'' value of
$g$ is not absolutely necessary. Indexing the quantities $\Delta_C$,
$\Delta_A$, $g$, $\kappa$, $\gamma$ would unnecessarily complicate the
notation without adding new physical effects. As a numerical example
we set $\gamma= 2\times 10^{7}$ s$^{-1}$, $\kappa=\gamma/2$,
$g=2.5\gamma$, $\Delta_C=-0.8 \kappa$, $\Delta_A=10^3 \gamma$ and
$k^{-1}=0.3 \mu$m. 

Having fixed the detunings, the only remaining control parameters are
the pump strengths $\eta_i$ ($i=r,b$) of the two modes and normalized
such that $|\eta_i|^2/(\Delta_C^2+\kappa^2)$ gives the stationary
photon number in the mode $i$ when no atoms are present. The intensity
of pumping is limited by the low saturation assumption for the atomic
transition, i.e the photon number in the fields must be well below the
saturation photon number of the atomic transitions,
$(\Delta_A^2+\gamma^2)/2 g^2$. In this limit we can associate two
independent dipoles with the two transitions. These atomic dipoles
being analogous to a microscopic dielectric medium shift the resonance
frequency of the interacting mode and incoherently scatter photons out
of this mode. The spatial dependence of the coupling between atom and
fields follows the exponential mode function. Hence the stationary
cavity photon number varies with the atomic position. For a
sufficiently slow atom the effect exerted by the nonresonant field can
be incorporated into an effective ``adiabatic'' optical potential.
This potential resulting from the red and blue fields is plotted in
figure 1 for the given parameter settings. The pumping constants have
been set to $\eta_r=1200$ and $\eta_b=1500$, giving rise to photon
numbers 8800 and 13700 for empty modes, respectively, and limiting
maximum saturation 10 \%. Although with these intensities the atom is
kept far enough from the surface (65nm), we take the conservative
Van der Waals attraction into account. Numerically we use
$U_{\mbox{\tiny VdW}}=5\times 10^{-3}/x^3$, characteristic of the
ground state of the ${}^{85}$Rb \cite{landragin96} which is a possible
candidate for the atomic system with its ${}^2\mbox{S}_{1/2}
\leftrightarrow 5{}^2\mbox{P}_{3/2}$, $5{}^2\mbox{P}_{1/2}$
transitions at 780 and 795 nm, respectively (the penetration constants
$k$ and $2 k$ can be then achieved with 48.3$^o$ and 63.5$^o$ incidence
angles in a n=1.45 refractive index medium).

Let us now qualitatively discuss the process of capturing and cooling
an atom.  Far from the surface an atom feels mainly an attractive
force due to the long range red detuned field.  Approaching the
surface, the atom shifts both mode resonances. The blue detuned mode
frequency is pushed to higher, the red one to lower values (the atom
is in the ground state). Since the detuning $\Delta_C$ is negative,
the photon number in the blue field decreases while it increases in
the red one. As a consequence the effective potential energy
decreases. However, the field amplitude variation cannot follow
instantaneously the atomic motion. This time delay due to the finite
response time $1/\kappa$ of the modes is the crucial reason for the
resulting nonconservative atomic motion. The potential energy decrease
comes to effect slightly after the bounce. On its way out, the atom is
subject to a weaker blue detuned field and a stronger red detuned
field as compared to its way in. Therefore, as outlined in the
introduction, in each atom-surface collision the atom loses a small
fraction of its kinetic energy. It cannot escape after the first
bounce and can finally be held for long periods in the trap. The
trapping time is limited only by extra heating processes (background
collisions) or spontaneous emission which, owing to the low saturation,
is relatively small.

Let us now turn to a more accurate simulation of the system based on a
semiclassical model established recently in ref.~\cite{domokos01}. A
central simplification in treating the interaction of the atomic
dipole with the field modes is the adiabatic elimination of the
internal atomic dynamics. This can safely be done if we assume large
atom-pump detuning, $\Delta_A >> g, \gamma$. The effect of the atom in
this limit can be regarded as a moving linearly polarizable particle
(atom or molecule) with dispersive and absorptive properties given by:
\begin{equation}
U_0= \frac{g^2 \Delta_A}{\Delta_A^2+\gamma^2}\; , \qquad 
\Gamma_0= \frac{g^2 \gamma}{\Delta_A^2+\gamma^2}\; .
\end{equation}
In this simple model the dipole effect is identical for the
red $r$ and blue $b$ fields. With the numerical example considered
here $U_0=0.0125 \kappa$, i.\ e., the atomic perturbation is very
small compared to the linewidth; hence the detuning $\Delta_C$ needed
to adjust the pumping frequency to a steep enough slope of the
spectral line of the cavity mode. The spontaneous scattering rate
$\Gamma_0 \approx 10^{-4}$ is negligible compared to other
frequencies of the system.

In a further approximation we treat the atomic center-of-mass motion
semiclassically. This is a good approximation as long as the atom
is well localized in both position and momentum space. It
can be fulfilled when $\hbar k^2 /M
\gamma < 1$. Of course the semiclassical treatment gets invalid if the
atom is cooled to the lowest few bound quantum states of the trap. In
ref.~\cite{domokos01} we showed that the consistent treatment of the
noise (momentum diffusion) requires additionally an approximate
quantum description of the field. In our case, this is very well
fulfilled since photon numbers on the order of $10^4$ are involved.

A set of coupled, stochastic differential equations can now be
established to simulate the dynamics:
\begin{eqnarray}
\dot x & = & \frac{p}{M}, \nonumber \\
\dot p & = & -\hbar U_0 \left( |\alpha_r|^2 -\frac{1}{2}\right)
         \nabla f_r^2(x) -\hbar U_0 \left( |\alpha_b|^2
         -\frac{1}{2}\right) \nabla f_b^2(x) + \xi_p, \nonumber \\
\dot \alpha_r & = & \eta_r + i \left(\Delta_C-U_0 f^2_r(x)\right)\alpha_r
                - \left(\kappa + \Gamma_0 f^2_r(x)\right)\alpha_r +
                \xi_{\alpha_r},\nonumber\\
\dot \alpha_b & = & \eta_b + i \left(\Delta_C-U_0 f^2_b(x)\right)\alpha_b
                - \left(\kappa + \Gamma_0 f^2_b(x)\right)\alpha_b +
                \xi_{\alpha_b},
\end{eqnarray}
where the variables $x$ and $p$ represent the position and the
momentum of the atomic center-of-mass, $\alpha_r$ and $\alpha_b$
denotes the complex field amplitudes of the red and blue fields,
respectively. 
The terms $\xi$ represent the noise and are defined via the diffusion
coefficients:
\begin{eqnarray}
\left\langle \xi^*_{\alpha_i} \xi_{\alpha_i} \right\rangle & = & 
\frac{1}{2}\left(\kappa+\Gamma_0 f_i^2(x) \right), \nonumber \\
\left\langle \xi_p \xi_p \right\rangle & = & \sum_{i=r,b} 2 \Gamma_0\left(|\alpha_i|^2-\frac{1}{2} \right)
   \left( (\hbar \nabla f_i(x))^2 + \hbar^2 k_i^2\bar{u^2}
   f_i^2(x)\right),
\end{eqnarray}
with $i=r,b$. While the first term of the momentum diffusion
coefficient represents dipole heating, the second term accounts for
momentum fluctuations due to the random direction of spontaneously
emitted photons. Here $k_i$ corresponds to the wavevector of the
transition quasiresonant with $i=r$ and $i=b$ fields. The factor $u^2$
is characteristic of the angular distribution of spontaneous
emission. Note that both terms are proportional to $\Gamma_0$ and
decrease for larger detunings. Surprisingly, the noise terms
associated with the momentum and the field amplitudes are correlated:
\begin{equation}
\left\langle \xi_{\alpha_i} \xi_p \right\rangle =  i \Gamma_0 \alpha_i
\hbar f_i(x) \nabla f_i(x) \; ,\quad\quad i=r, b \; .
\end{equation}

\begin{figure}
\oneimage[width=14cm]{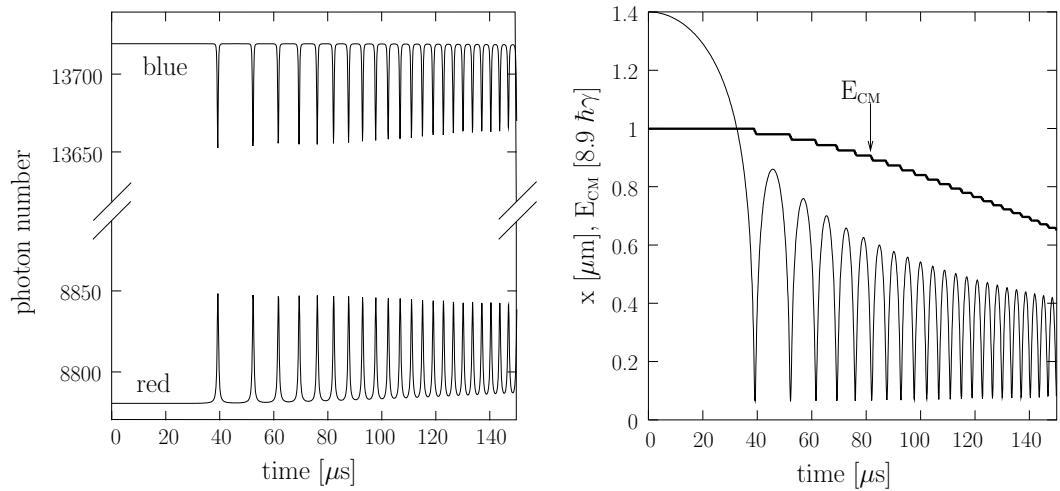}
\caption{Left: Variation of the photon numbers in the blue and red
detuned modes. Right: Oscillations in the trap with decreasing
amplitude. Bold line represents the mechanical energy of the atom
including the kinetic energy and the adiabatic potential energy. The
scale for this curve assumes the units of the trap depth that is about
8.9 $\hbar\gamma$.}
\label{f.2}
\end{figure}
It is instructive to first solve these equations without noise to gain
some insight into the cooling mechanism resulting from the coupled
atom-field dynamics. In figure 2 we plot the time evolution of
important quantities in the dynamics.  The photon numbers in the red
and blue detuned fields exhibit sharp anticorrelated peaks.  The most
drastic changes happen at the time duration when the atom is close to
the inner turning points and feels the strongest acceleration, while
the time evolution is smoother at the outer turning points.  After a
bounce the photon numbers return relatively fast to the steady--state
value.  Then the atomic motion becomes approximately conservative in
the potential presented in figure 1. The bold line in figure 2 (right
side) represents the motional energy associated with the atomic
center-of-mass motion (kinetic and potential energy).  We clearly see
a stepwise decrease of the energy at each inelastic reflection, in
agreement with our previous qualitative explanation. For this plot the
units are scaled by the depth of the effective potential 8.9
$\hbar\gamma$. The energy loss in a single bounce is about 0.17
$\hbar\gamma$, which also indicates how large can the initial velocity
be (about 7 cm/s) such that the cooling mechanism still leads to
trapping after a single bounce.

\begin{figure}
\onefigure[width=14cm]{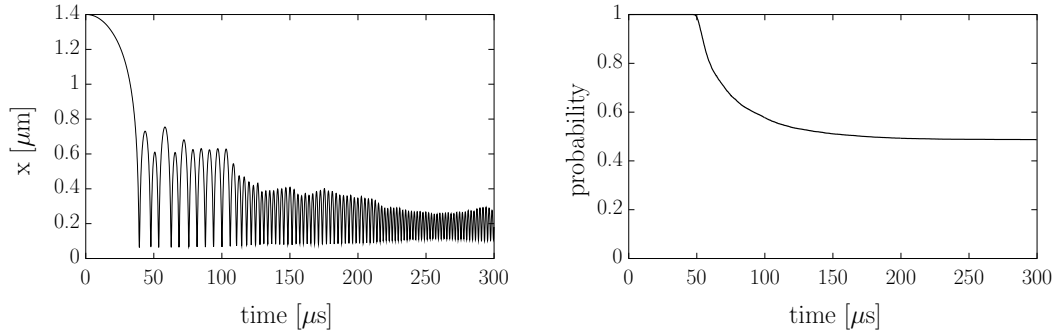}
\caption{Left: Typical trajectory of a trapped atom. Right: The probability of trapping the atom.}
\label{f.3}
\end{figure}
Figure 3 presents the time evolution including noise sources. In the
left graph we selected a trajectory of an atom trapped for a long
period. The evolution of the position resembles to the result plotted
in figure 2, however, the amplitude of the oscillations takes up some
random element. As the oscillations are cooled down, the atom is
localized relatively far from the strong field region and the
spontaneous photon scattering rate decreases.  In some cases the noise
causes the atom to leave after the first or eventually after several
bounces. Hence a very important question is the long-time trapping
probability.  Figure 3 (right side) shows this trapping probability of
an atom as a function of time, averaged over a large number N=10000
trajectories.  The initial rapid decrease can be attributed to the
particles that come too close to the surface in the first bounce or
leave the trap due to an incoherent random recoil event during the
first cycle. After only very few bouncing times the trapping
probability tends to a finite value of 48.5\% for the given
parameters. Hence an atom which survives the first few bounces will be
trapped for a very long time.  This provides for a very efficient
feeding mechanism of our dynamic evanescent-wave trap. Even on much
longer time scales the further decrease of the trapping probability
practically vanishes.
\begin{figure}
\onefigure[width=7cm]{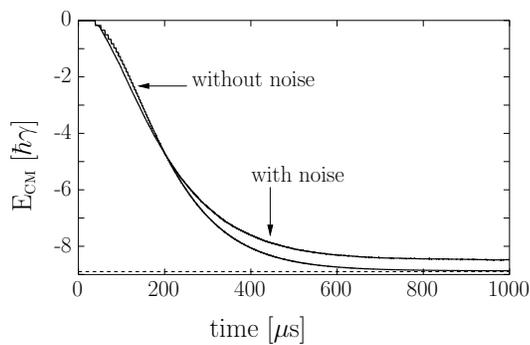}
\caption{The mechanical energy associated with the atomic
center-of-mass motion. The result for the noisy dynamics involves an
averaging over 10000 trajectories, however, only those are counted
that correspond to a trapped atom (48.5 \% of the trajectories finally).
Dashed line indicates the potential minimum value. }
\label{f.4}
\end{figure}
Figure 4 shows the evolution of the kinetic energy of the atom on a
millisecond range. The results from the simulations including and
excluding noise have been plotted (this latter stepwise curve is the
same as in figure 2). They predict quite similar damping of the atomic
kinetic energy. However, in the noiseless case the final state is an
atom at rest in the potential minimum, while the noise induces some
residual oscillations. In this latter case, the asymptotic kinetic
energy corresponds to $\gamma/2.5 \approx\kappa$, which is in good
agreement with the results obtained in other cavity configurations
\cite{domokos01}.  At such a low final temperature, the probability of
escaping from the trap is indeed extremely low. In this regime, the
actual trapping time might be limited by other mechanisms, such as for
example background collisions and technical noise of the
apparatus. Also note that, the oscillation frequency in the adiabatic
trap being $1.71 \times 10^6$ s$^{-1}$, the final temperature is
equivalent to very few (about 4-5) vibrational excitations in the
trap. That is, the cooling mechanism is able to drive the atom into
the quantized motion regime. Then the semiclassical description is no
longer valid and one must resort to a full quantum model involving
spatially extended atomic states in the trap.

In conclusion, we calculated the nonconservative motion of an atom in
a dynamic optical trap. The probability of capturing an atom from free
space has been found to be 48.5 \%. The dynamically varying fields can
effectively damp the atomic oscillations in the trap, which yields
finally very long trapping times limited by technical noise in the
setup. A next important step is of course to identify a specific
realization of the scheme, and to perform a detailed 3D simulation to
calculate the expected performances.

 
\acknowledgments

We thank for the discussions with Peter Horak and Markus Gangl. This
work was supported by the Austrian Science Foundation FWF (Project
P13435, Z30-TPH Wittgenstein Preis). P.~D.~ acknowledges the financial
support by the National Scientific Fund of Hungary under contracts
No. T023777 and F032341.

\end{document}